\documentclass[a4paper]{jpconf}
\usepackage{graphicx}
\usepackage{amsmath,amssymb,amscd}
\usepackage{multirow}
\usepackage{comment}

\usepackage{slashed}
\usepackage{tikz}

\newcommand{\be}{\begin{equation} }
\newcommand{\ee}{\end{equation}}
\newcommand{\cL}{\mathcal{L}}
\newcommand{\MSb}{$\overline{\text{MS}}$}

\usetikzlibrary{decorations.markings,snakes}
\tikzset{
  fermionline/.style={line width=1pt,postaction={decorate},
    decoration={markings,
      mark=at position 0.5 with {\draw[-stealth] (0,0)--(2pt,0);}}},
  bosonline/.style={line width=1pt,decorate,
    decoration={snake,amplitude=1,segment length=4}},
  higgsline/.style={line width=1pt,dashed}
}

\begin{document}
\title{Higgs inflation and vacuum stability}

\author{Javier Rubio}

\address{Institut de Th\'{e}orie des Ph\'{e}nom\`{e}nes Physiques, 
\'{E}cole Polytechnique F\'{e}d\'{e}rale de Lausanne, CH-1015 Lausanne, 
Switzerland}

\ead{javier.rubio@epfl.ch}

\begin{abstract}
Inflation is nowadays a well-established paradigm consistent with all the observations.  The precise nature of the inflaton is however unknown and its role could be played by any candidate able to imitate a scalar condensate in the slow-roll regime. 
The discovery of a fundamental scalar in the LHC provides the less speculative candidate. Could the Higgs field itself be responsible for inflation? Do we really need to advocate new physics to explain the properties of the Universe at large scales? Which is the relation between the Standard Model parameters and the inflationary observables? What happens if our vacuum becomes unstable below the scale of inflation?  We present an overview of Higgs inflation trying to provide answers to the previous questions with special emphasis on the vacuum stability issue.\end{abstract}

\section{The Higgs boson in the sky}
\vspace{3mm}

Nature seems to be extremely simple. The only outcome of LHC experiments till date is a scalar boson with a mass of $125-126$ GeV and properties that closely resemble those of the Standard Model Higgs \cite{Aad:2012tfa,Chatrchyan:2012ufa}.  No sign of new physics beyond the Standard Mode has appeared whatsoever. On top of that, the Planck results favor the simplest realization of the inflationary scenario: a single field model of inflation in which no significant amount of non-gaussianities or isocurvature perturbations are produced \cite{Ade:2015lrj}.

Given the minimalistic scenario presented above, it is certainly tempting to try to identify the scalar field found in the LHC with that responsible for the flatness, homogeneity and isotropy of the Universe at large scales. Unfortunately, the self-coupling of the Higgs field highly exceeds the value needed to generate the right amount of primordial density perturbations \cite{Linde:1983gd}. A possible way out is the inclusion of a non-minimal coupling\footnote{The existence of this coupling is indeed required by renormalization \cite{Birrell:1982ix}.} to gravity with no further modifications of the Standard Model Lagrangian. This is the starting point of Higgs-inflation \cite{Bezrukov:2007ep}. The relevant part of the Lagrangian density reads
 \be
\label{lagr}
\frac{\cal L}{\sqrt{-g}}= \frac{M_P^2+ \xi h^2}{2}   R -\frac{1}{2}\left(\partial h\right)^2- \frac{\lambda}{4}(h^2-v^2)^2\,,
\ee
with $M_P$ the Planck mass, $h$ the Higgs field in the unitary gauge and $v$ its vacuum expectation value. The non-minimal coupling $\xi$ is a free parameter that cannot be fixed from the model itself and must be determined by observations.
Note that, for sufficiently large values of the Higgs field (namely $h\gg M_P/\sqrt{\xi}$), the theory becomes approximately scale invariant. This asymptotic symmetry will play a central role in the further developments\footnote{The promotion to an exact symmetry of the theory and its phenomenological consequences were considered in Refs. \cite{Shaposhnikov:2008xb,GarciaBellido:2011de,Bezrukov:2012hx,Rubio:2014wta,GarciaBellido:2012zu}.}. 

\begin{figure}
\centering
  \includegraphics[scale=0.27]{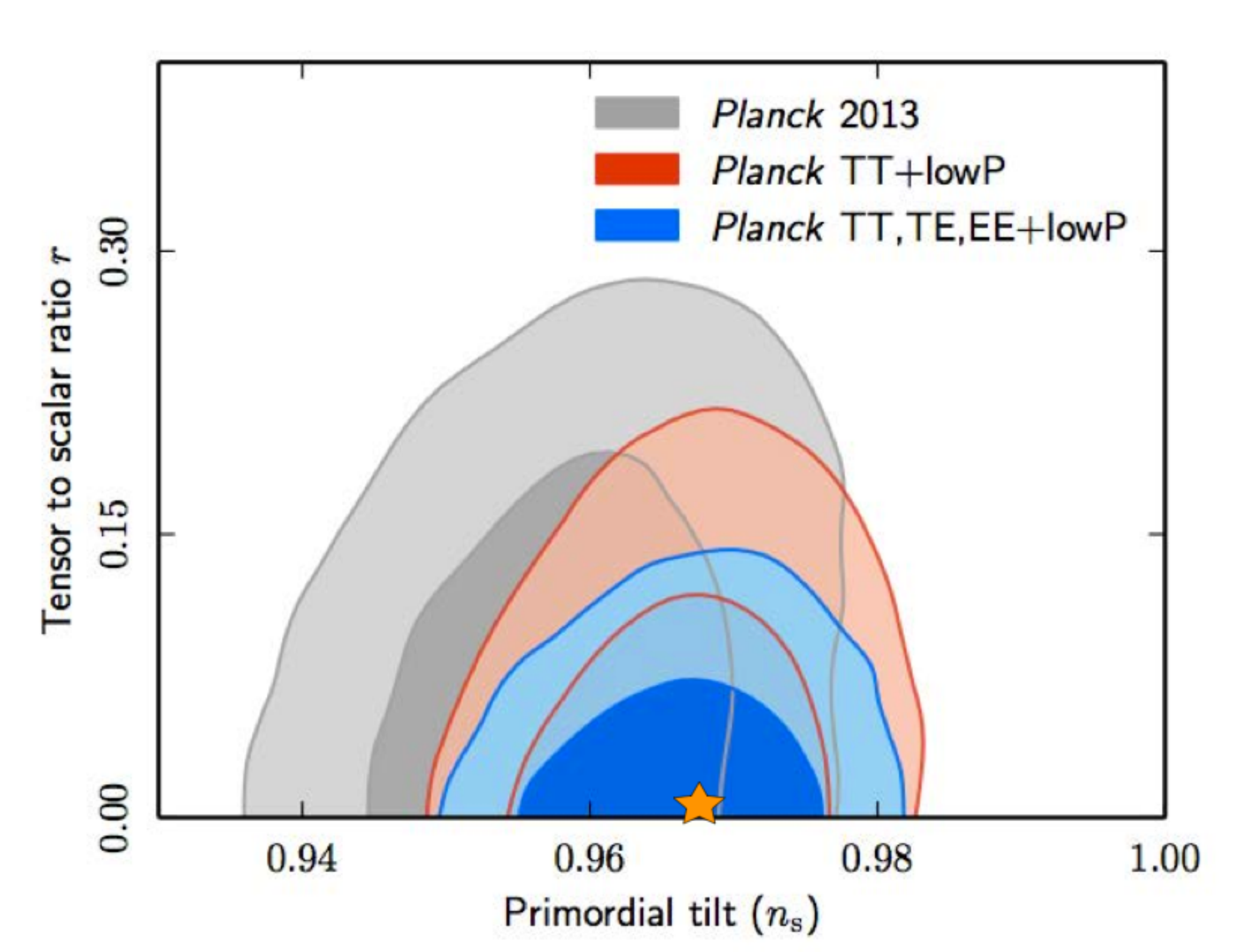}
\caption{Comparison of the tree level predictions of Higgs inflation (yellow star) and the Planck data \cite{Ade:2015lrj}.}\label{planckfig}
\end{figure}

The analysis of inflation is more easily performed in the so-called Einstein frame, in which the gravitational part of the action takes the usual Einstein-Hilbert form. This is obtained by performing 
a conformal transformation 
 \begin{equation}\label{conformal}
g_{\mu\nu}\rightarrow \Omega^2 g_{\mu\nu}\,,\hspace{5mm}\textrm{with}\hspace{5mm}\Omega^2 = 1+\frac{\xi h^2}{M_P^2}~,
 \end{equation}
together with a field redefinition
  \begin{equation}\label{connection}
\frac{d\chi}{d h}=\sqrt{\frac{\Omega^2+6\xi^2h^2/M_P^2}{\Omega^4}}\hspace{5mm} \longrightarrow \hspace{5mm}  
\chi\simeq
\begin{cases} 
   h\,, & h\ll \frac{M_P}{\xi}\,,\\
\sqrt{\frac{3}{2}} M_P \log \Omega^2(h)\,, & h\gg \frac{M_P}{\xi}\,,
   \end{cases}
\end{equation}
 to make the kinetic term of the Higgs field canonical. All the non-linearities in the original frame are moved to the Higgs potential, which after the field redefinition, becomes non-polynomial (and therefore non-renormalizable)\footnote{We omit the exponentially small correction proportional to the vacuum expectation value of the Higgs field in the expression for the potential at large field values.}
 \begin{equation}\label{Vtree}
 V(\chi)=
\begin{cases} 
      \frac{\lambda}{4}(\chi^2-v^2)^2\,, & \chi\ll \frac{M_P}{\xi}\,, \\
 \frac{\lambda M_P^4}{4\xi^2}\left(1-e^{-\sqrt{2/3}\chi/M_P}\right)^2\,, & \chi\gg \frac{M_P}{\xi}\,.
   \end{cases}
 \end{equation}
 The asymptotic shift-symmetry of the previous expression at large field values is the Einstein-frame manifestation of the approximate scale invariance we started with. With the standard initial chaotic conditions, the flattening potential \eqref{Vtree} allows for inflation.
 
 The analysis of inflation follows the usual slow-roll approach. The COBE normalization fixes the amplitude of the potential, $ \lambda/\xi^2\simeq 4\times 10^{-11}$.  For $\lambda\sim {\cal O}(1)$, the non-minimal coupling $\xi$ is required to be significantly large, $\xi \sim {\cal O}(10^4)$, but still much smaller than the value able to produce noticeable effects at low energies \cite{Bezrukov:2007ep}.
 The model predicts a spectral tilt for the primordial scalar perturbations around $n_s\simeq 0.97$ and a small tensor-to-scalar ratio $r\simeq 0.003$. As shown in Fig.~\ref{planckfig}, both numbers are fully compatible with the bounds obtained by the Planck collaboration  \cite{Ade:2015lrj} and very close to the predictions of the Starobinsky model of inflation\footnote{The small differences are associated to the different preheating mechanisms in the two models \cite{Bezrukov:2011gp}.}   
   \cite{Starobinsky:1980te,Mukhanov:1981xt}. Since we are dealing with a single field model of inflation with canonical kinetic terms, neither isocurvature perturbations nor non-Gaussianities are produced.
\section{A potential problem: vacuum stability}
\vspace{3mm}
The previous predictions of Higgs inflation are based on the classical action and do not take into account possible changes in the shape of the inflationary potential due to quantum effects. One of the main points to consider is the running of the coupling constants and, in particular, the logarithmic running of  the Higgs self-coupling $\lambda$. 

One should be tempted to think that the precise value of the Higgs coupling at the inflationary scale should not have any effect on the inflationary observables since the non-minimal coupling $\xi$ can be tuned at will in order to satisfy the normalization condition $\lambda/\xi^2\simeq 4\times 10^{-11}$. However, the situation turns out to be much more complicated due to the particular value of the Higgs mass.

\begin{figure}[]
\centering
\begin{minipage}[b]{0.48\linewidth}
  \includegraphics[scale=0.95]{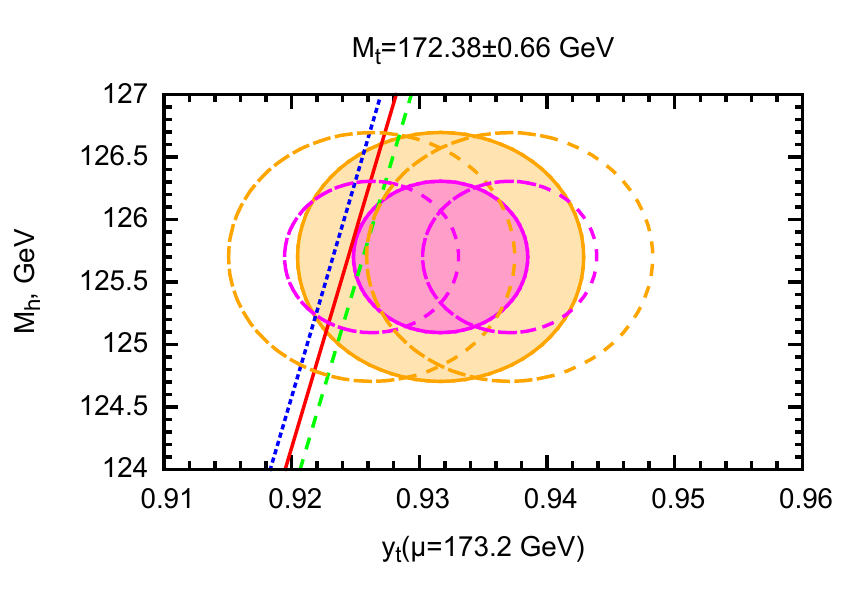}
\end{minipage}
\quad
\begin{minipage}[b]{0.45\linewidth}
  \includegraphics[scale=0.85]{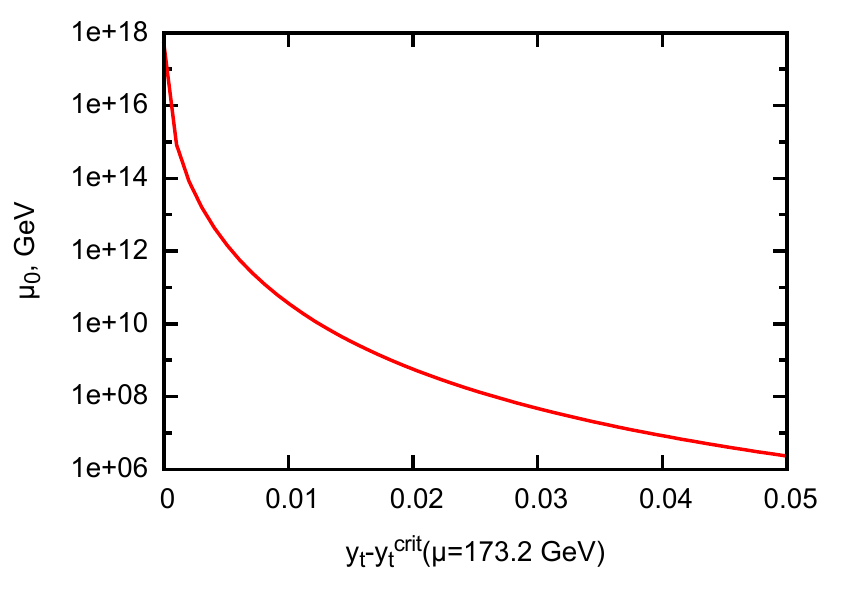}
\end{minipage}
\caption{\textit{Left}: Stability/Metastability regions of the Standard Model in the \MSb \,\,scheme  at $\mu=173.2$ GeV \cite{Bezrukov:2014ina}. The red diagonal line stands for the critical value of the top Yukawa coupling $y_t^{\rm crit}(M_h)$ leading to a negative self-coupling $\lambda$ at a scale $\mu_0$ below the Planck scale (dashed lines accounts for the uncertainty in the strong coupling constant $\alpha_s$). The Standard Model is absolutely stable to the left of these lines and metastable to the right.  The filled ellipses correspond to the experimental values of $y_t$ extracted from the latest CMS determination of the Monte-Carlo top quark mass $M_t = 172.38 \pm 0.10\, {\rm (stat)} \pm 0.65\, {\rm(syst)}\, {\rm GeV} $, if this is identified with the pole mass  \cite{CMS}. 
The dashed ellipses encode the shifts associated to the ambiguous relation between pole and Monte Carlo masses. \textit{Right}: Energy scale $\mu_0$ at which the Higgs self-coupling $\lambda$ becomes negative as a function of the deviation of the top Yukawa coupling $y_t$ from the critical value $y_t^{\rm crit}$ \cite{Bezrukov:2014ina}.}\label{figCMS}
\end{figure}

Among the many possible values that the Higgs mass could have taken, Nature has chosen one that allows to extend the Standard Model all the way up till the Planck scale while staying in the perturbative regime. The scalar self-coupling decreases with energy to reach a minimum at energies close to $M_P$ and starts increasing thereafter. Whether it remains positive till the Planck scale or becomes negative at a smaller scale $\mu_0$ depends on the precise values of the top Yukawa coupling and the Higgs mass (cf. Refs.~\cite{Bezrukov:2012sa,Degrassi:2012ry,Buttazzo:2013uya,Bezrukov:2014ina} for details). Although the latest CMS data \cite{CMS} are perfectly consistent with the absolute stability of the Standard Model within the experimental and theoretical uncertainties (cf. Fig.~\ref{figCMS}), one should not exclude the possibility that other experiments may be able to establish the metastability of the electroweak vacuum in the future. Should the successful Higgs inflation idea be abandoned in that case? Answering this question is the purpose of the next sections.

\section{Higgs inflation as an effective field theory}
\vspace{3mm}
The detailed analyses about vacuum stability performed in Refs.~\cite{Bezrukov:2012sa,Degrassi:2012ry,Buttazzo:2013uya} are based on flat space considerations and should not be extrapolated to the case in which gravitational effects are included since issues such as the determination of the lifetime of the Universe strongly depend on the details of the ultraviolet completion~\cite{Branchina:2013jra,Branchina:2014usa,Branchina:2014rva}. 

The presence of gravity makes Higgs inflation non-renormalizable. Quantum corrections should be introduced by interpreting the theory as an effective field theory in which a particular set of higher dimensional operators are included. But which set of operators? 

It seems clear that we cannot add arbitrary operators in the Einstein frame, since this would automatically spoil the flatness of the potential. In the absence of an ultraviolet completion for the Standard Model non-minimally coupled gravity, the choice of higher dimensional operators can be only based on the self-consistency of the procedure  \cite{Bezrukov:2010jz}. To have a (partially) controllable link between the low and high energy parameters of the model, we will only add those counterterms that are required to make the theory finite at every order in perturbation theory, i.e. those with the structure generated by the original Lagrangian via radiative corrections.
 The finite parts associated to these counter terms are assumed to be small and to have the hierarchy of the loops producing them (i.e. the coefficients in front of the operators coming from two-loop diagrams are much smaller than those coming from one-loop diagrams, etc\ldots). Finally, the procedure will be required to maintain the symmetries of the original theory, and, in particular the approximate scale invariance of Eq.~\eqref{lagr} at large field values. This can be accomplished by using dimensional regularization\footnote{This regularization automatically discards all power- law divergences.} with an asymptotically scale-free t'Hooft-Veltman parameter\footnote{This is a Jordan-frame definition.}
\begin{equation}\label{muI}
\mu^2 \propto M_P^2/2  + \xi H^\dagger H\,.
\end{equation}

\section{From the Standard Model to the Chiral Standard Model}
\vspace{3mm}

\begin{figure}
\centering
\begin{minipage}[b]{0.45\linewidth}
  \includegraphics[scale=0.37]{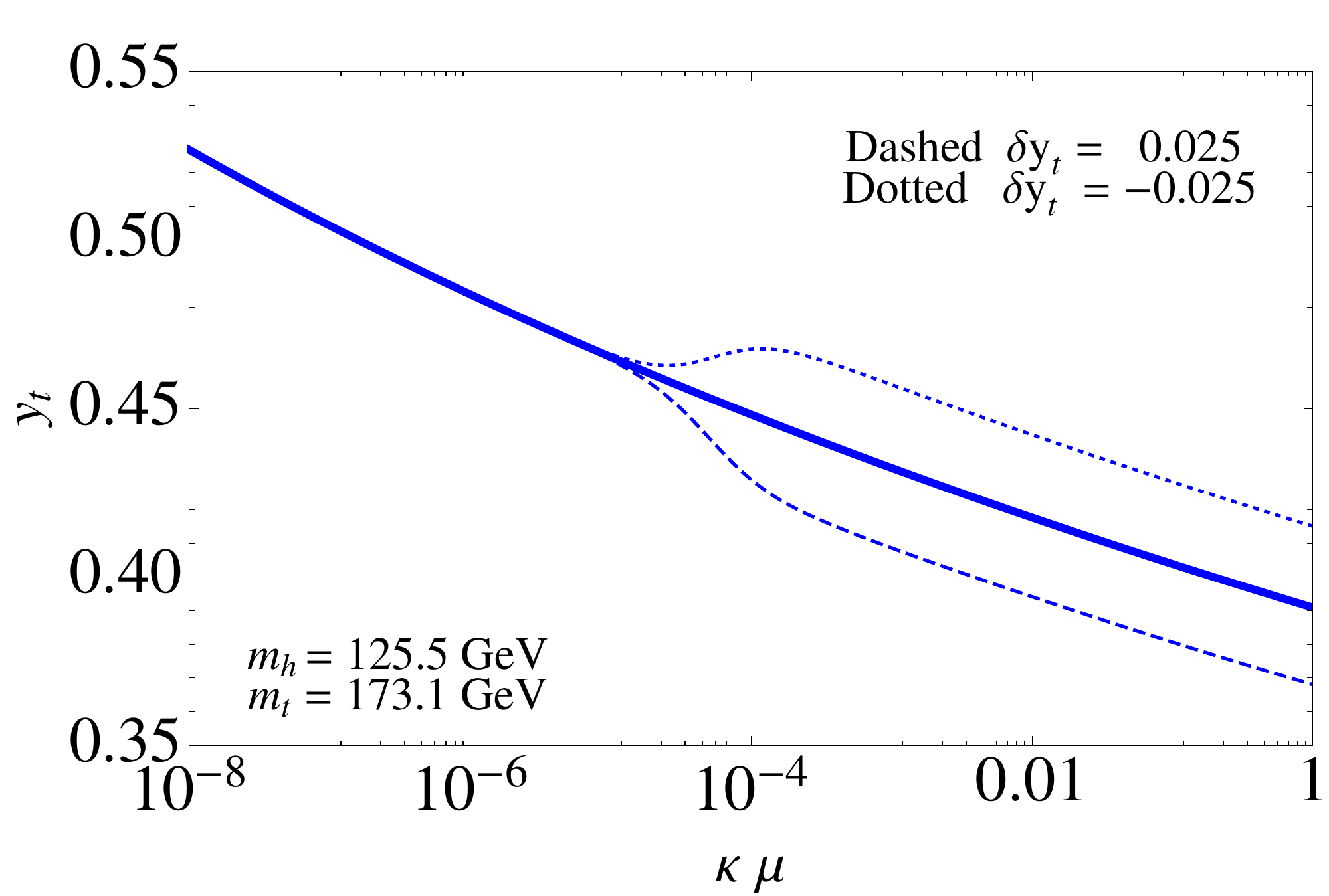}
\end{minipage}
\quad
\begin{minipage}[b]{0.45\linewidth}
  \includegraphics[scale=0.37]{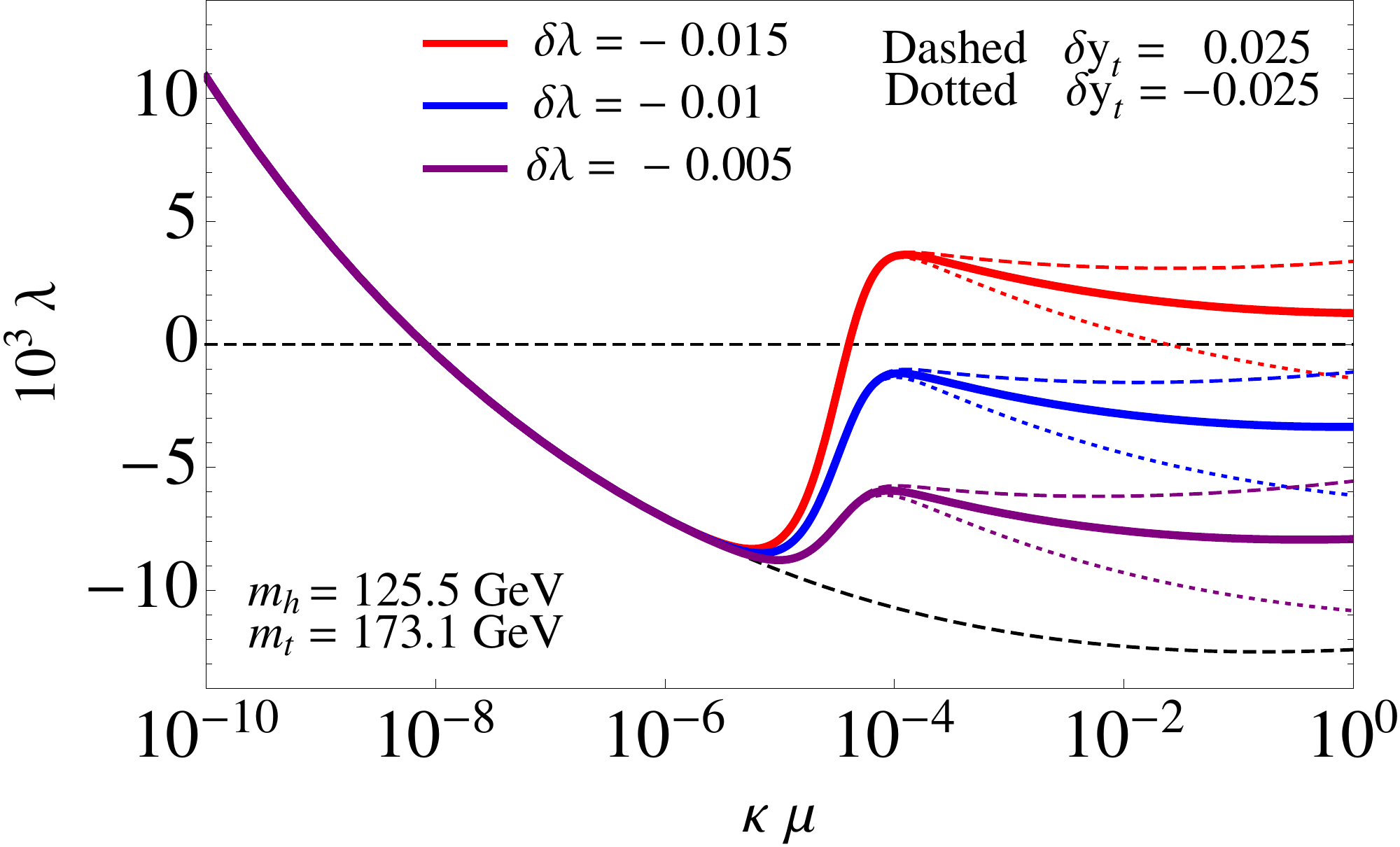}
\end{minipage}
\caption{Renormalization group evolution of  the top Yukawa coupling $y_t$ and the Higgs self-coupling $\lambda$ as a function of the field-dependent renormalization scale $\mu(\chi)$ in Planck units ($\kappa=M_P^{-1}$). The effect of the jumps is localized at a given energy scale and does not significantly modify the running of the couplings in the asymptotic low and high energy regions.}\label{fig:jumps}
\end{figure}
 
 Given the self-consistent framework presented in the previous section, which is the relation between the Standard Model parameters and the inflationary observables? To answer this question consider the propagation of a given fermion, for instance the top quark\footnote{This will give the most relevant contribution.}, in the background of the Higgs field. When written in the Einstein frame, the linear Yukawa interaction between these two species becomes modified by a conformal factor
\begin{equation}
  \label{Ltree}
  \cL_F = i\bar\psi_t\slashed\partial\psi_t + \frac{y_t}{\sqrt{2}} F(\chi)\bar\psi_t\psi_t\,, \hspace{10mm}\textrm{with} \hspace{10mm}F(\chi) \equiv  \frac{h(\chi)}{\Omega(\chi)}\,.
\end{equation}
The divergences associated to the 1-loop diagrams 

\begin{equation}
  \tikz[baseline=-0.5ex] {
    \draw[fermionline] (-3em,0) -- (3em,0);
    \draw[higgsline] (-1.5em,0) arc (180:0:1.5em);
    \draw[fill] (-1.5em,0) circle (0.2em) node[below] {$yF'$};
    \draw[fill] (1.5em,0) circle (0.2em) node[below] {$yF'$};
  }
  +
  \tikz[baseline=-0.5ex] {
    \draw[fermionline] (-3em,0) -- (0,0);
    \draw[fermionline] (0,0) -- (3em,0);
    \draw[higgsline] (0,0) arc (-90:270:1.5em);
    \draw[fill] (0,0) circle (0.2em) node[below] {$yF''$};
  }\,,
\end{equation}
are eliminated, as usual, by adding counterterms with the proper coefficients $a_1$ and $a_2$ in $1/\bar\epsilon$ and arbitrary finite parts $\delta y_{t1}$ and $\delta y _{t2}$
\begin{eqnarray}\label{count1}
  \delta\cL_\text{ct} &\sim&
  \left(a_1\frac{y_t^3}{\bar\epsilon}+\delta y_{t1}\right) F'^2F \bar\psi\psi
  +
  \left(a_2\frac{y_t\lambda}{\bar\epsilon}+\delta y_{t2}\right) F''(F^4)''\bar\psi\psi \,.
\end{eqnarray}
Here $1/\bar\epsilon$ stands for $1/\epsilon-\gamma+\ln4\pi$, with $\gamma=0.5772$ the Euler-Mascheroni constant, and the primes denote derivatives with respect to $\chi$. The structure of the new terms differs from that in Eq.~\eqref{Ltree}, as expected, since we are dealing with a non-renormalizable theory.

At small field values ($ \chi\ll M_P/\xi$), the conformal factor $\Omega(h)$ equals one, and the function $F(\chi)$ in Eqs.~\eqref{Ltree} and \eqref{count1} becomes linear in the field ($F=\chi$ and $F'=1$).  The Higgs-fermion interaction in the Einstein frame reduces then to the standard (renormalizable) Yukawa interaction and the finite part $y_{t1}$ can be reabsorbed in the definition of the top Yukawa coupling, as in any renormalizable theory. This allows us to use the usual Standard Model renormalization group equations to relate the values of the couplings at the electroweak scale with those at $\chi=M_P/\xi$. What happens at higher energies? When approaching $\chi\sim M_P/\xi$, the function $F(\chi)$ becomes 
\begin{equation}
F(\chi) =
      \frac{M_P}{\sqrt{\xi}}\left(1-\e^{-\sqrt{2/3}\chi/M_p}\right)^{1/2}\,,
\end{equation}
and the previously existing counterterms become exponentially suppressed.  This gives rise to an effective jump of the top Yukawa coupling at that scale\footnote{We neglect the running of $\delta y_{t1}$.} \cite{Bezrukov:2014ipa}
\begin{equation}
  \label{ytjump}
  y_t(\mu) \to y_t(\mu)+ \delta y_t\left[
    F'^2-1
  \right]\,.
\end{equation}
Beyond that point (in particular for $\chi>M_P$), the function $F(\chi)$ becomes rapidly a constant, $F(\chi)\approx   \frac{M_P}{\sqrt{\xi}}$,
and the Higgs field decouples from the Standard Model particles. Higgs inflation becomes a chiral Standard Model \cite{Longhitano:1980iz} in which the radial component of the Higgs field has been integrated out. 
 
Threshold effects as those giving rise to \eqref{ytjump} appear also in the Higgs sector. The evaluation of the bubble diagrams 
\begin{eqnarray}
  \label{1loopVscalar0}
  \tikz[baseline=-0.5ex]{\draw[higgsline] (0,0) circle (1.4em);}
  & =& \frac{1}{2}\Tr\ln\left[\Box-\left(\frac{\lambda}{4}(F^4)''\right)^2\right]\,,  \hspace{10mm}
   \label{1loopVfermion0}
  \tikz[baseline=-0.5ex]{\draw[fermionline] (0,-1.5em) arc (270:-90:1.4em);}
   = -\Tr\ln\left[i\slashed\partial+y_tF\right] \,,
  \end{eqnarray}
involved in the computation of the effective potential, and the addition of the associated counterterms,  produces a rapid change in the self-coupling of the Higgs field at\footnote{Again, we neglect the running of the finite part $\delta \lambda$.} 
 $\chi\sim M_P/\xi$ \cite{Bezrukov:2014ipa}
 \begin{equation}
  \label{lambdajump}
  \lambda(\mu) \to \lambda(\mu)
  + \delta\lambda \left[
    \left(F'^2+\frac{1}{3}F''F\right)^2-1
    \right]\,.
\end{equation}
As shown in Fig.~\ref{fig:jumps}, the effect of the jumps  \eqref{ytjump} and \eqref{lambdajump} is localized at $\chi\sim M_P/\xi$ and does not significantly modify the running of the couplings in the asymptotic low and high energy regions.

\section{Restoring the asymptotics}
\begin{figure}
\centering
\begin{minipage}[b]{0.47\linewidth}
  \includegraphics[scale=0.4]{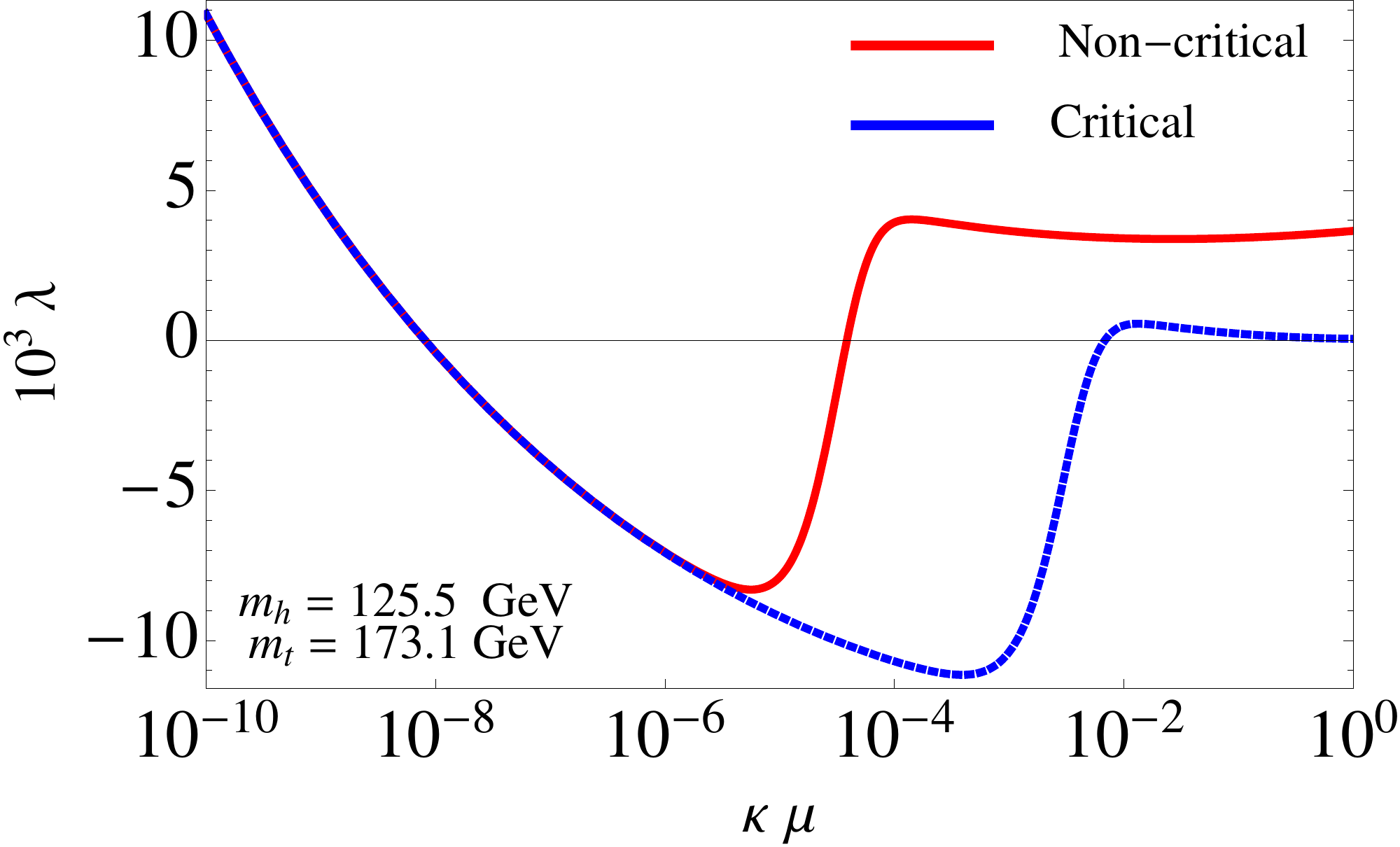}
\end{minipage}
\quad
\begin{minipage}[b]{0.49\linewidth}
\begin{tikzpicture}
\draw[<->] (0,4) node[left] {$V$} -- (0,-0.24) -- (7.5,-0.24) node[below] {$\chi$};
\draw[thick,blue]
(0,0) .. controls (0.1,-0.3) and (0.2,-0.3) ..
(0.4,0) .. controls (0.7,0.6) and (1.1,0.6) ..
(1.4,0) .. controls (1.8,-1) and (2,-1.5) ..
(2.7,-1.5) .. controls (3.5,-1.5) and (3.9,-0.7) ..
(4,0) .. controls (4.2,1) and (4,4) ..
(7.5,4);
\node[below] at (0.25,-0.25) {$\scriptstyle v_{\text{EW}}$};
\node[below,xshift=-0.4em] at (1.5,-0.25) {$\mu_0$};
\node[below,xshift=0.7em] at (4,-0.25) {$\frac{M_P}{\xi}$};
\node[below] at (5.7,-0.25) {$M_P$};
\end{tikzpicture}

\end{minipage}
\caption{\textit{Left}: Jump-based restoration of a positive self-coupling $\lambda$ at the inflationary scale in the case of a metastable vacuum. The red and blue lines correspond respectively to the non-critical  ($\xi=1500$, $\delta y_t=0.025$, $\delta\lambda=-0.015$) and critical ($\xi=15$, $\delta y_t=0$, $\delta\lambda=-0.0133$) cases discussed in the text. \textit{Right}: Sketch of the associated effective inflationary potential.}\label{potentialR}
\end{figure}
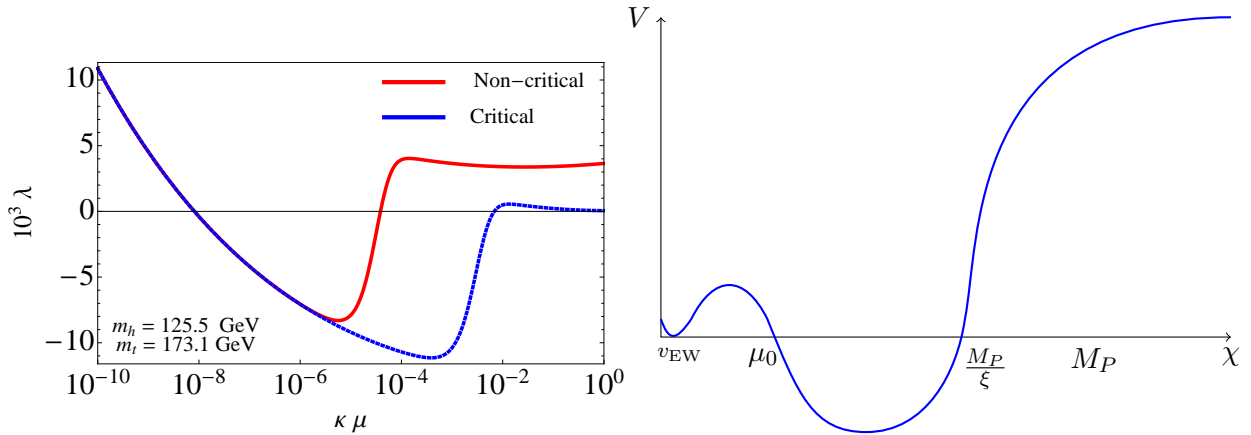

\vspace{3mm}
The precise value of the finite parts  $\delta y_t$  and $\delta \lambda$ in Eqs. \eqref{ytjump} and $\eqref{lambdajump}$ is unknown. They depend on the particular ultraviolet completion of  Higgs inflation and cannot be determined from the effective field theory itself. 

If the jumps   $\delta \lambda$ and $\delta y_t$ are much smaller than the corresponding coupling constants at the inflationary scale --i.e. if  $\delta\lambda \ll \lambda(M_P/\xi)$ and $\delta y_t \ll y_t(M_P/\xi)$ -- then Higgs inflation requires the absolute stability of the vacuum and provides a clear connection between the Higgs and top masses and the inflationary observables \cite{Bezrukov:2009db}.

Note however that the smallness of the self-coupling $\lambda$ at the inflationary scale appears as the consequence of a non-trivial cancellation between the bosonic and fermionic degrees of freedom in the Standard Model. A situation in which $\delta\lambda \sim \lambda$, $\delta y_t \sim y_t$ should not be,  \textit{a priori}, excluded. If that turns out to be the case,  the previously discussed connection between low and high energy observables is lost but, at the same time, the model allows for inflation even if our vacuum is not completely stable \cite{Bezrukov:2014ipa}. 

As shown in Fig.~\ref{potentialR}, a small jump $\delta\lambda\sim 10^{-2}$ is enough to convert a negative self-coupling constant below $M_P/\xi$ into a positive one above that scale. The potential in this case displays two minima. The first one (the shallowest and narrowest one) corresponds to the usual electroweak vacuum. The second one (the deepest and widest one) lays between the scale $\mu_0$ at which the self-coupling becomes negative and the transition scale $M_P/\xi$ in which the jumps in the coupling constants appear. Beyond $\chi\sim M_P$, the potential becomes asymptotically flat  and allows for inflation with the standard chaotic initial conditions.  

\section{Non-critical vs. critical Higgs inflation}
\vspace{3mm}
The fate of Universe strongly depends on the interplay between the energy density at the end of inflation, the depth of the deeper minimum and the efficiency of the reheating process. If the backreaction of the particles created during preheating is large enough to modify the shape of the effective potential and to make the wrong vacuum disappear, the Higgs field will be able to roll down to the electroweak minimum. If this is not the case, the Higgs field will end in the minimum at Planckian values and the Universe will eventually collapse. 

\begin{figure}
\centering
\includegraphics[scale=0.5  ]{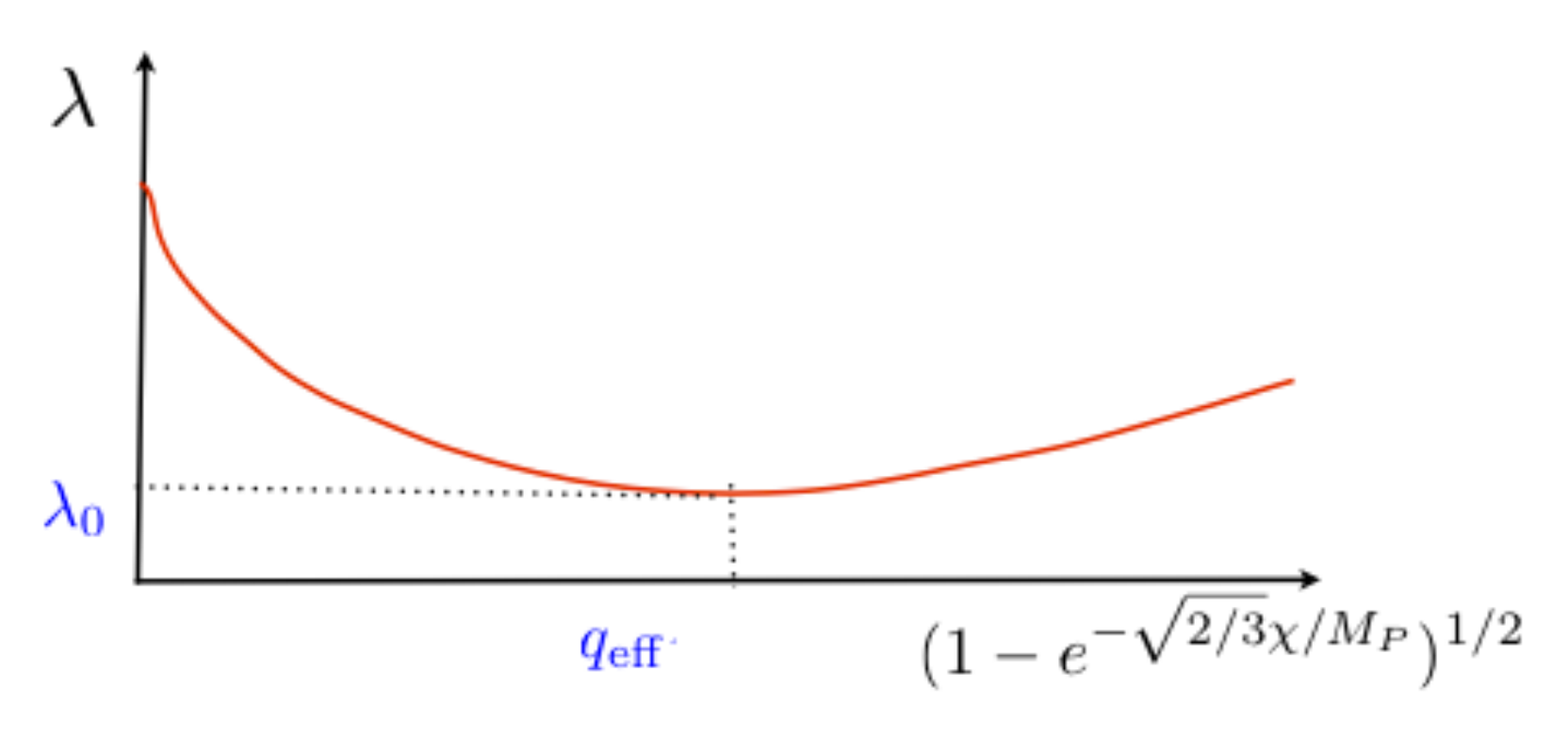}
\caption{Asymptotic behavior of the Higgs self-coupling around the inflationary scale as a function of the scale $\mu(\chi)$. }
\label{fig:runcrit}
\end{figure}

\begin{figure}
\centering
  \includegraphics[scale=0.36]{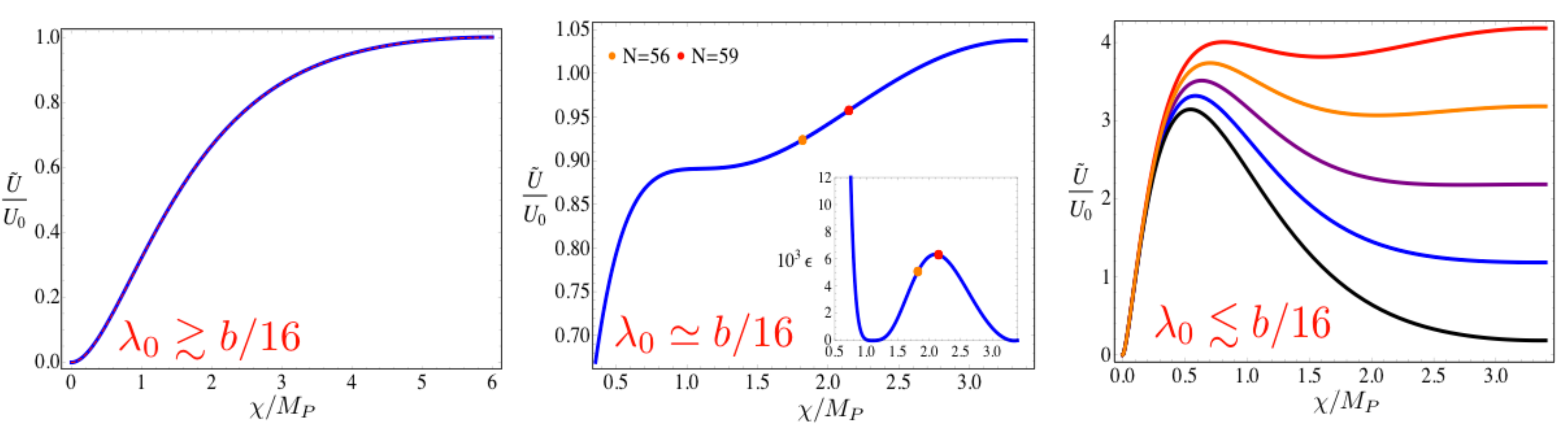}
\caption{Shape of the inflationary potential in the non-critical (left), critical (center) and forbidden (right) regimes. The behavior of the non-monotonic slow-roll parameter $\epsilon$ in the critical case is also shown.}\label{critfig}
\end{figure}

The relation between the energy density available at the end of inflation and the depth of the wrong minimum depends on the asymptotic behavior of the Higgs self-coupling at the inflationary scale. The last one can be parametrized as (cf. Fig.~\ref{fig:runcrit})
\begin{equation}
\label{Lrun}
\lambda(\chi) =\lambda_0+b \log^2 \left(\frac{\sqrt{1-e^{-\alpha\kappa\chi}}}{q_{\rm eff}}\right)\,,
\end{equation}
with $b\simeq 2.3 \times 10^{-5}$  $\alpha=\sqrt{2/3}$, $\kappa=M_P^{-1}$  and $\lambda_0$ and $q_{\rm eff}$ some functions of the top quark pole mass, the Higgs mass and the strong coupling constant \textit{at the inflationary scale} (i.e. after taking into account the rapid changes in the coupling constants at $\chi\sim M_P/\xi$), cf. Ref.~\cite{Bezrukov:2014bra} for details. The expression \eqref{Lrun} reveals the existence of three different regimes (cf. Fig.~\ref{critfig}):
\begin{itemize} 
\item {\it Non-critical regime}: For $\lambda_0 \gg b/16$, the form of the potential is almost independent of 
the precise values of the parameters appearing within the logarithmic correction. The potential effectively depends on the combination $\lambda/\xi^2$, which, as in the tree-level case, can be fixed by the COBE normalization. The non-minimal coupling $\xi$ associated to a situation like the one presented in Fig.~\ref{potentialR} ($\lambda\sim 10^{-3}$) turns out to be of order $\xi\sim 10^3$ \cite{Bezrukov:2009db}.  
\item{\it Critical regime}: If $\lambda_0\simeq b/16$, the second derivative of the potential is equal to zero at some intermediate field value between the beginning and the end of inflation and the first derivative becomes very small (but non-zero). The dependence of the running function $\lambda(\chi)$ on the parameters $\lambda_0$, $\xi$ 
and $q_{\rm eff}$ is now essential. The shape of the inflationary potential strongly differs from the tree-level case: it contains a very flat region around the inflection point in which most of the e-folds of inflation take place and a temporally increasing slope at larger field values. The associated slow-roll parameter is non-monotonic and allows to obtain a sizable tensor-to-scalar ratio \cite{Bezrukov:2014bra,Hamada:2014iga}, whose precise value strongly depends on the non-minimal coupling $\xi$, which is generically required to be small ($\xi\sim10$).

\item {\it Forbidden regime}: If $\lambda_0 \lesssim b/16$, the potential develops a wiggle and inflation is not longer possible.
\end{itemize}
 
\section{Preheating and symmetry restoration}
\vspace{3mm}

One of the main differences between non-critical and critical Higgs inflation is the value of the non-minimal coupling $\xi$. This parameter controls not only the energy scale of inflation, but also the energy scale $M_P/\xi$ at which the jumps in the coupling constants take place and the depth of the vacuum at large field values (cf. Fig.~\ref{potentialR}) . 

\subsection{Non-critical Higgs inflation with metastable vacuum}
\vspace{3mm}

\begin{figure}
\centering
\begin{minipage}[b]{0.47\linewidth}
  \includegraphics[scale=0.33]{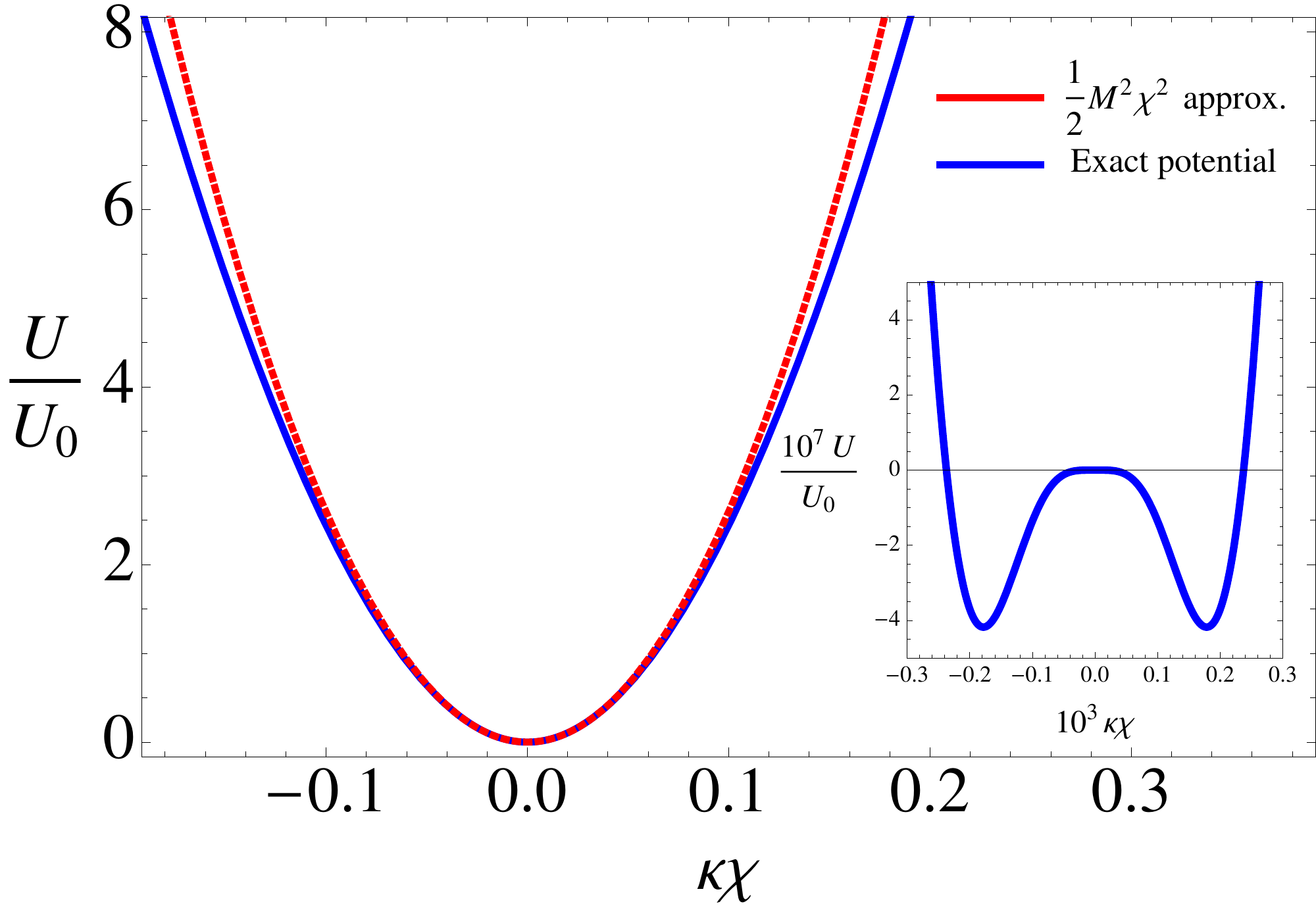}
\end{minipage}
\quad
\begin{minipage}[b]{0.47\linewidth}
  \includegraphics[scale=0.33]{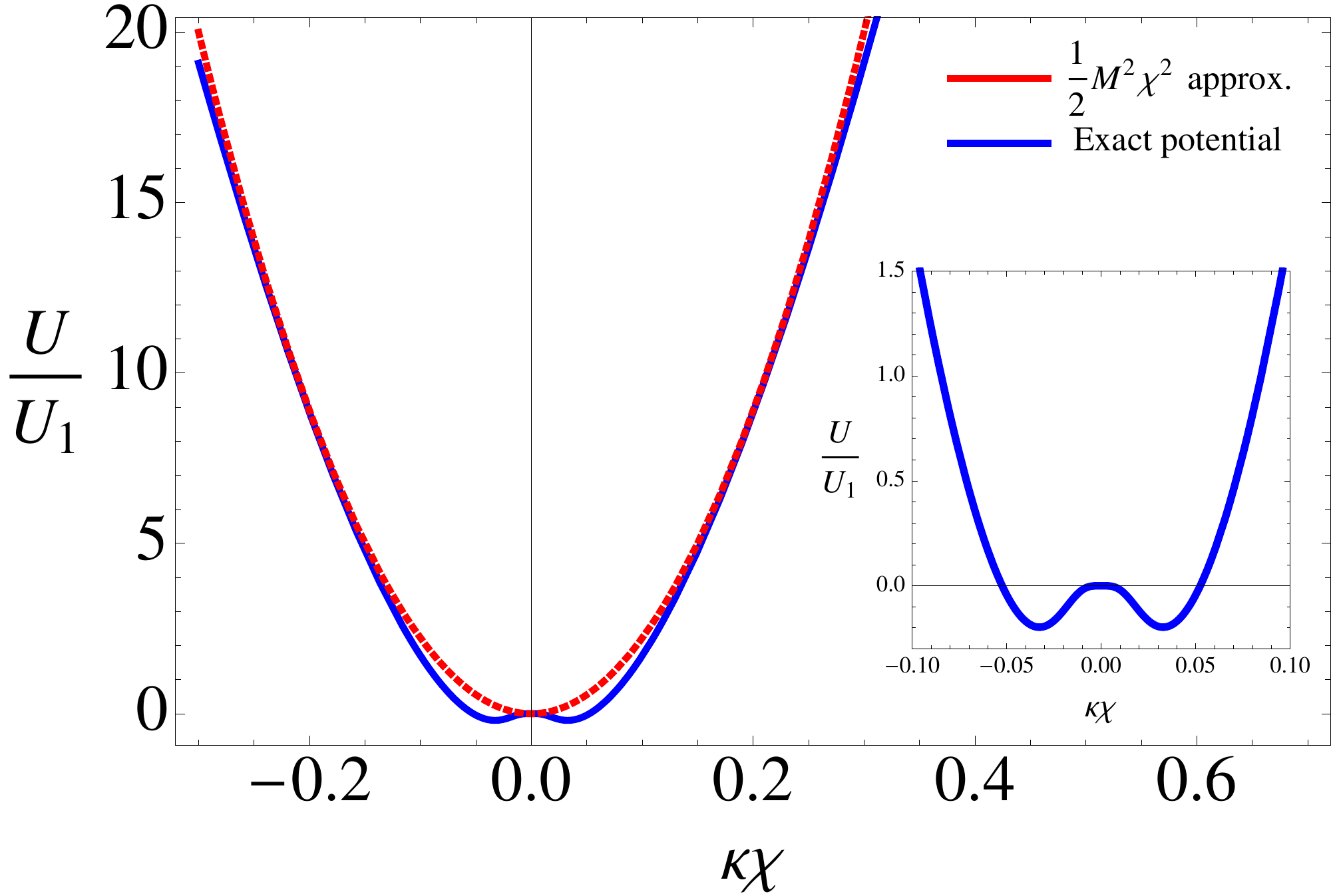}
\end{minipage}
\caption{\textit{Left}: Renormalization group enhanced potential in the non-critical case compared to a quadratic approximation $\frac{1}{2}M^2\chi^2$ with $M^2=\frac{\lambda M_P^2}{3\xi^2}$. The normalization scale $U_0$ is taken to be $U_0=(10^{-3} M_P)^4$.  \textit{Right}: Similar plot for the critical case. The normalization scale $U_1$ is given by $U_1=10^{-9} M_P^4$.}\label{Ecomp}
\end{figure}

\begin{figure}
\centering
\begin{minipage}[b]{0.47\linewidth}
  \includegraphics[scale=0.37]{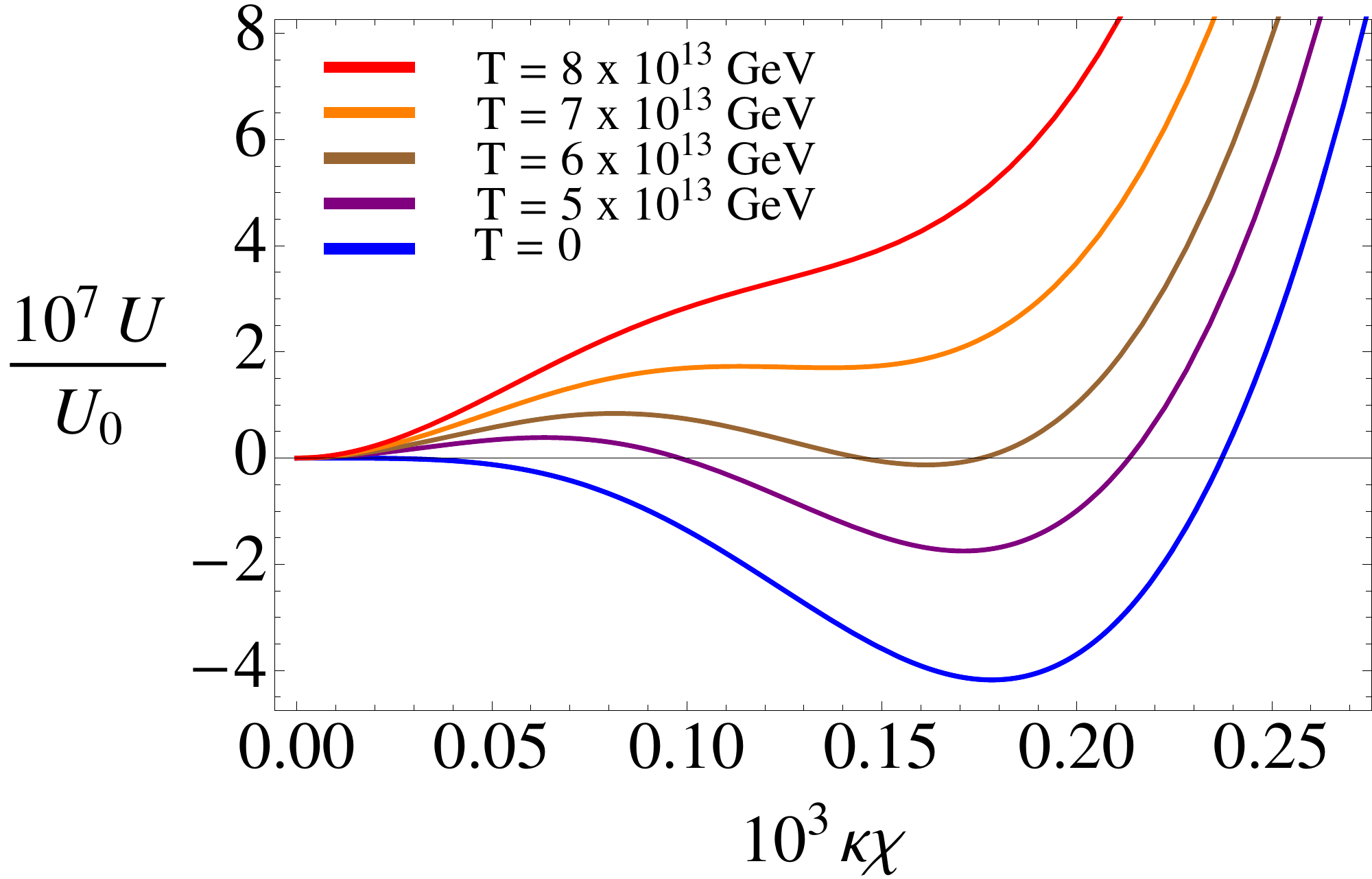}
\end{minipage}
\quad
\begin{minipage}[b]{0.47\linewidth}
  \includegraphics[scale=0.35]{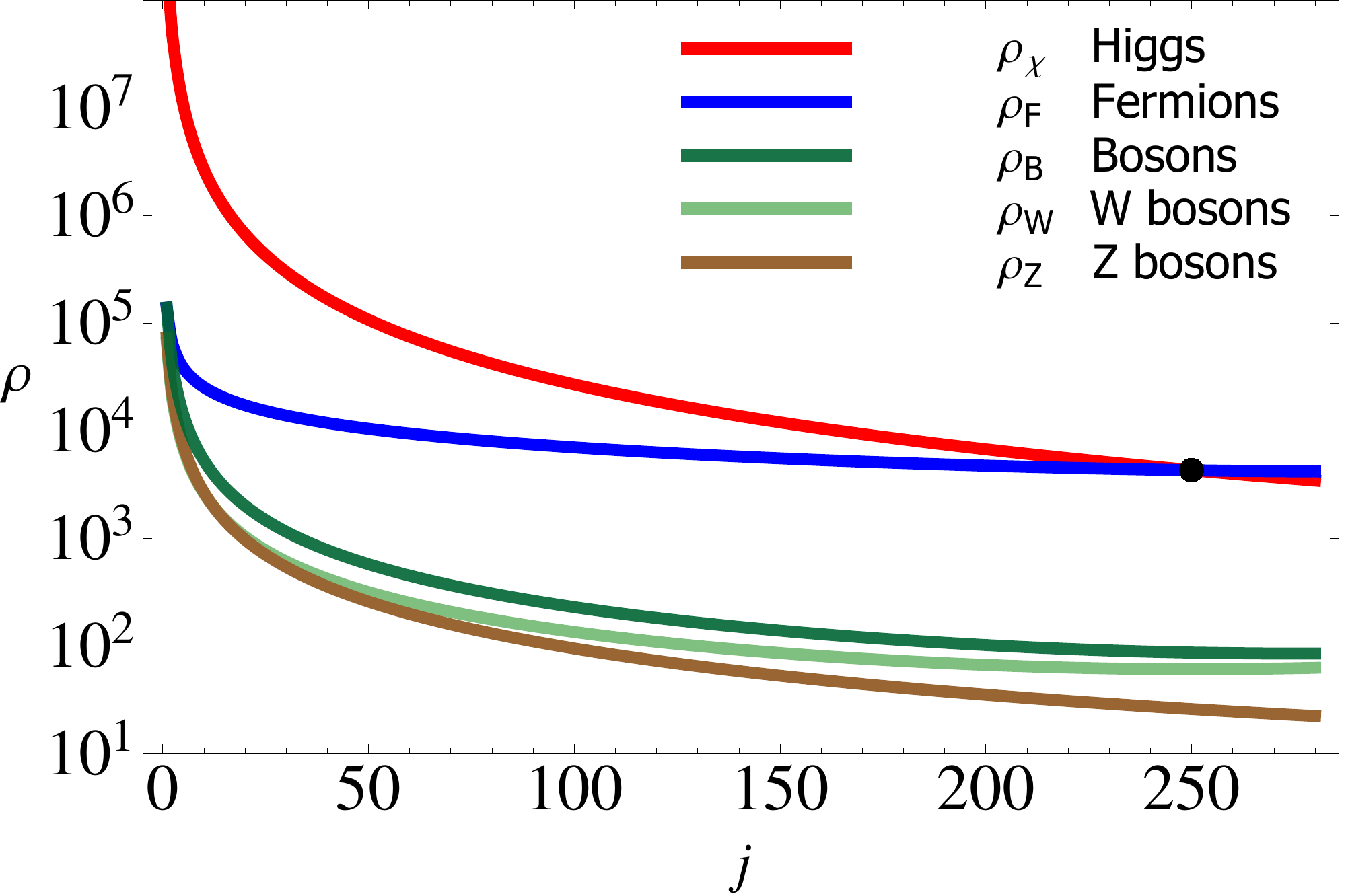}
\end{minipage}
\caption{\textit{Left}: High temperature effective potential for non-critical Higgs inflation ($m_H=125.5$\,GeV, $m_T= 173.1$\,GeV, $\xi=1500$, $\delta y_t = 0.025$, $\delta\lambda = -0.0153$). \textit{Right}: Evolution of the different energy densities (in $M^4$ units) for the non-critical case ($\lambda=3.4\times 10^{-3}$, $\xi=1500$) as a function of the number of semioscillations $j$ of the Higgs field around $\chi=0$.}\label{NCpreh}
\end{figure}

As shown in the left panel of Fig.~\ref{Ecomp}, the energy available at the end of inflation in the non-critical case ($\xi\sim10^3$) is much larger (by a factor $\sim10^7$) than the depth of the wrong vacuum. A detailed analysis of the particle production after the end of inflation \cite{Bezrukov:2014ipa} via the so-called Combined Preheating formalism  \cite{GarciaBellido:2008ab,Bezrukov:2008ut} reveals that the reheating temperature $T^{NC}_{RH}$ exceeds the restoration temperature $T^{NC}_+$ at which the extra minimum at large field values disappears (cf. Fig.~\ref{NCpreh})
\begin{equation}\label{TcritNC}
T^{NC}_{RH}> T^{NC}_+\,,\hspace{10mm}\textrm{with}\hspace{10mm}T^{NC}_{RH}\simeq 1.8 \times 10^{14}\hspace{2mm}\text{GeV}\,,\hspace{5mm} \textrm{and} \hspace{5mm} T^{NC}_+\simeq7\times10^{13}\,\,\,\text{GeV}\,.
\end{equation}
The Higgs field will then be able to relax to the Standard Model vacuum. In the subsequent evolution of the Universe the temperature will decrease and the second minimum will eventually reappear. The probability of decaying via tunneling through the barrier separating the two vacua turns out to be rather small and the lifetime of the electroweak vacuum highly exceeds the life of the Universe \cite{Anderson:1990aa,Arnold:1991cv,Espinosa:1995se,Espinosa:2007qp}. \textit{Non-critical Higgs inflation can take place with a graceful exit even if the Standard Model vacuum is not completely stable}.

\begin{figure}
\centering
\begin{minipage}[b]{0.47\linewidth}
  \includegraphics[scale=0.37]{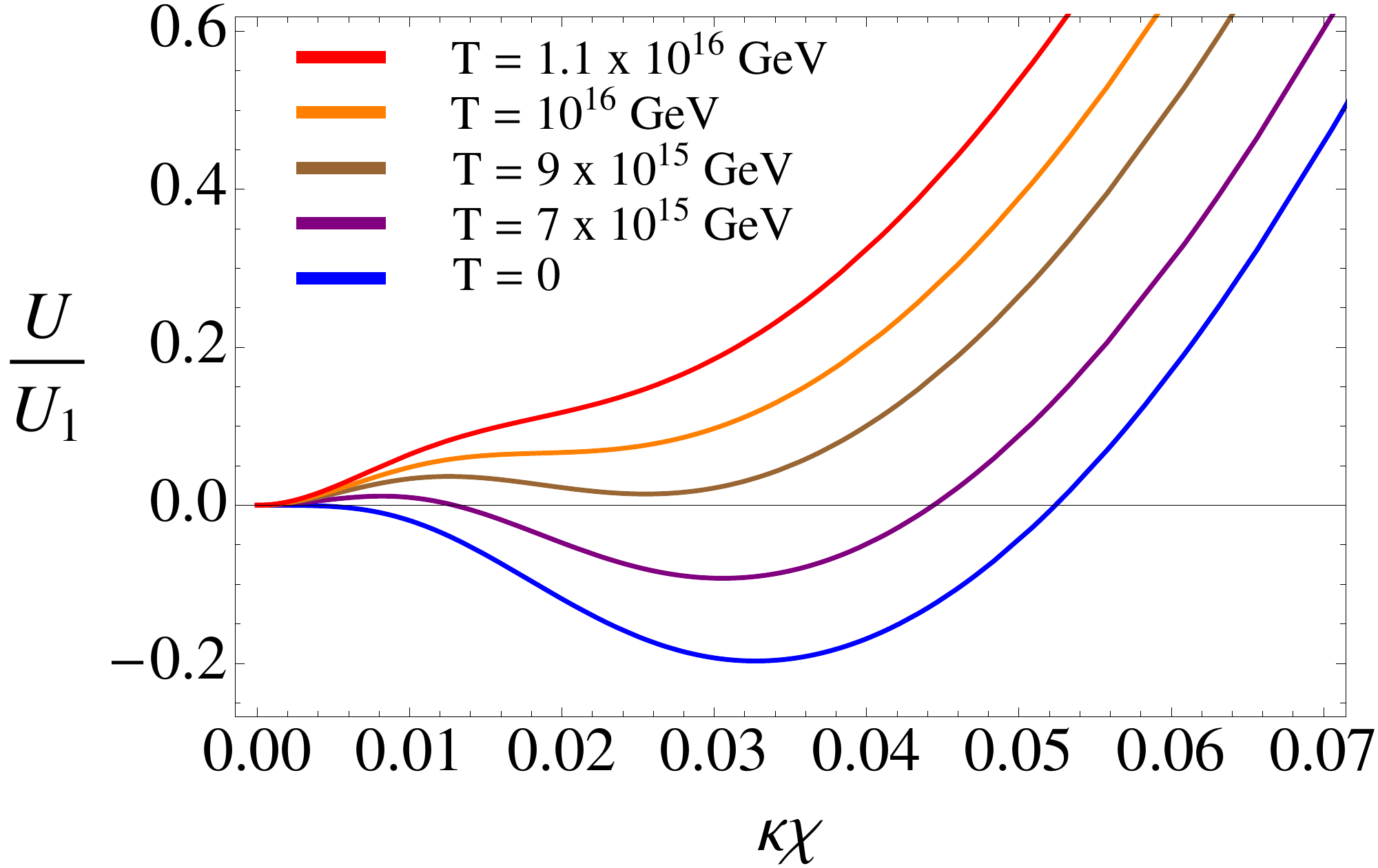}
\end{minipage}
\quad
\begin{minipage}[b]{0.47\linewidth}
  \includegraphics[scale=0.35]{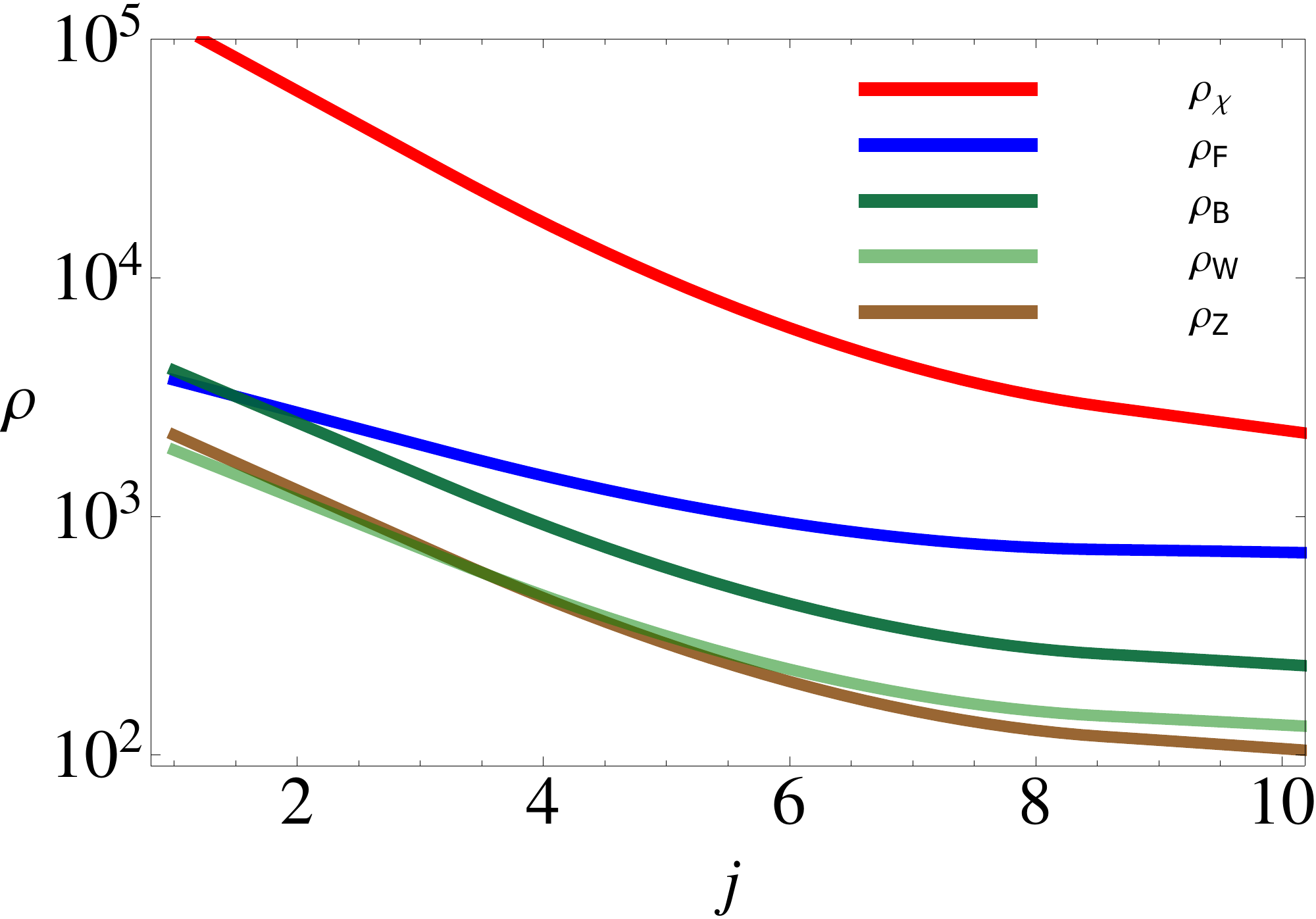}
\end{minipage}
\caption{\textit{Left}: High temperature effective potential for critical Higgs inflation ($m_H=125.5$\,GeV, $m_T= 173.1$\,GeV, $\xi=15$, $\delta y_t = 0$, $\delta\lambda = -0.01325$).  \textit{Right}: Evolution of the different energy densities (in $M^4$ units) for the critical case ($\lambda=3\times 10^{-4}$, $\xi=15$) as a function of the number of semioscillations $j$ of the Higgs field around $\chi=0$. }
\end{figure}

\subsection{Critical Higgs inflation with metastable vacuum}
\vspace{3mm}
The situation in the critical case ($\xi\sim10$) is completely different. The energy stored in the Higgs field after inflation is {\em comparable} to the height of the barrier separating the two vacua (cf. the right panel of Fig.~\ref{Ecomp}). An upper bound on the reheating temperature $T^C_\text{RH}$ can be obtained by simply converting the energy density of the inflaton field at the end of inflation ($V^{1/4}\simeq 6\times 10^{16}$~GeV) into an instantaneous radiation temperature. For the number of degrees of freedom produced during Combined Preheating, this temperature turns to be smaller than the restoration temperature $T_+^{C}$ 
\begin{equation}
T^{C}_{RH}< T^{C}_+\,,\hspace{10mm}\textrm{with}\hspace{10mm}T^C_\text{RH}< 7.85 g_*^{-1/4} \times 10^{16}\,\, \text{GeV}\,, \hspace{5mm} \textrm{and} \hspace{5mm}  T^C_+\simeq10^{16}\,\,\,\text{GeV}\,.
\end{equation}
The shape of the potential in the non-critical case remains unchanged after preheating and the system inevitably relaxes towards the deep minimum at Planckian values. When the energy on the field becomes equal to the amplitude of the barrier, the expansion of the Universe stops and the scale factor begins to shrink till the Universe collapses. \textit{Critical Higgs inflation does necessarily require the absolute stability of the Standard Model vacuum}.

\section{Conclusions}
\vspace{3mm}
The value of the Higgs mass together with the apparent absence of new physics at the LHC allows to speculate with the possibility of consistently extending the Standard Model all the way up till the Planck scale. The Higgs field itself can be responsible for inflation if a minimalistic, and at the same time compelling, non-minimal coupling to gravity is allowed. No new particles beyond the electroweak scale are required.
The non-minimal coupling makes the Standard Model non-renormalizable, which requires the addition of an infinite number of counterterms. The absence of an ultraviolet completion introduces some uncertainties associated to the unknown finite parts of the counterterms. If these are small compared to the values of the couplings at the inflationary scale, Higgs inflation requires the absolute stability of the Standard Model vacuum. On the other hand, if the finite parts are comparable to the couplings at the inflationary scale, (non-critical) Higgs inflation can take place even for the case of a metastable vacuum. The associated predictions  ($n_s=0.97$, $r=0.003$) are universal and  in excellent agreement with the latest CMB results.

\section*{Acknowledgments}
\vspace{3mm}
It is a pleasure to thank F. Bezrukov, M. Shaposhnikov and G. Karananas for our collaborations and discussions. This work was partially supported  by the Swiss National Science Foundation.


\section*{References}
\begin{thebibliography}{99}  

\bibitem{Aad:2012tfa}
  G.~Aad {\it et al.}  [ATLAS Collaboration],
  Phys.\ Lett.\ B {\bf 716} (2012) 1
  [arXiv:1207.7214 [hep-ex]].
  
\bibitem{Chatrchyan:2012ufa}
  S.~Chatrchyan {\it et al.}  [CMS Collaboration],
  Phys.\ Lett.\ B {\bf 716} (2012) 30
  [arXiv:1207.72 35 [hep-ex]].


\bibitem{Ade:2015lrj}
  P.~A.~R.~Ade {\it et al.}  [Planck Collaboration],
  arXiv:1502.02114 [astro-ph.CO].
  
\bibitem{Linde:1983gd}
  A.~D.~Linde,
  Phys.\ Lett.\ B {\bf 129} (1983) 177.
  
\bibitem{Birrell:1982ix}
  N.~D.~Birrell and P.~C.~W.~Davies,
  
\bibitem{Bezrukov:2007ep}
  F.~L.~Bezrukov and M.~Shaposhnikov,
  Phys.\ Lett.\ B {\bf 659} (2008) 703
  [arXiv:0710.3755 [hep-th]].
  
  
\bibitem{Shaposhnikov:2008xb}
  M.~Shaposhnikov and D.~Zenhausern,
  Phys.\ Lett.\ B {\bf 671} (2009) 187
  [arXiv:0809.3395 [hep-th]].
      
\bibitem{GarciaBellido:2011de}
  J.~Garcia-Bellido, J.~Rubio, M.~Shaposhnikov and D.~Zenhausern,
  Phys.\ Rev.\ D {\bf 84} (2011) 123504
  [arXiv:1107.2163 [hep-ph]].
  
\bibitem{Bezrukov:2012hx}
  F.~Bezrukov, G.~K.~Karananas, J.~Rubio and M.~Shaposhnikov,
  Phys.\ Rev.\ D {\bf 87} (2013) 9,  096001
  [arXiv:1212.4148 [hep-ph]].
  
\bibitem{Rubio:2014wta}
  J.~Rubio and M.~Shaposhnikov,
  Phys.\ Rev.\ D {\bf 90} (2014) 2,  027307
  [arXiv:1406.5182 [hep-ph]].
      
\bibitem{GarciaBellido:2012zu}
  J.~Garcia-Bellido, J.~Rubio and M.~Shaposhnikov,
  Phys.\ Lett.\ B {\bf 718} (2012) 507
  [arXiv:1209.2119 [hep-ph]].
  
\bibitem{Bezrukov:2011gp}
  F.~L.~Bezrukov and D.~S.~Gorbunov,
  Phys.\ Lett.\ B {\bf 713} (2012) 365
  [arXiv:1111.4397 [hep-ph]].
  
  
\bibitem{Starobinsky:1980te}
  A.~A.~Starobinsky,
  Phys.\ Lett.\ B {\bf 91} (1980) 99.
  
\bibitem{Mukhanov:1981xt}
  V.~F.~Mukhanov and G.~V.~Chibisov,
  JETP Lett.\  {\bf 33} (1981) 532
   [Pisma Zh.\ Eksp.\ Teor.\ Fiz.\  {\bf 33} (1981) 549].
 
  
\bibitem{Bezrukov:2012sa}
  F.~Bezrukov, M.~Y.~Kalmykov, B.~A.~Kniehl and M.~Shaposhnikov,
  JHEP {\bf 1210} (2012) 140
  [arXiv:1205.2893 [hep-ph]].
  
\bibitem{Degrassi:2012ry}
  G.~Degrassi, S.~Di Vita, J.~Elias-Miro, J.~R.~Espinosa, G.~F.~Giudice, G.~Isidori and A.~Strumia,
  JHEP {\bf 1208} (2012) 098
  [arXiv:1205.6497 [hep-ph]].
  
\bibitem{Buttazzo:2013uya}
  D.~Buttazzo, G.~Degrassi, P.~P.~Giardino, G.~F.~Giudice, F.~Sala, A.~Salvio and A.~Strumia,
  JHEP {\bf 1312} (2013) 089
  [arXiv:1307.3536 [hep-ph]].
  
\bibitem{Bezrukov:2014ina}
  F.~Bezrukov and M.~Shaposhnikov,
  arXiv:1411.1923 [hep-ph].
  
  \bibitem{CMS}
https://cds.cern.ch/record/1951019?ln=en


\bibitem{Branchina:2013jra}
  V.~Branchina and E.~Messina,
  Phys.\ Rev.\ Lett.\  {\bf 111} (2013) 241801
  [arXiv:1307.5193 [hep-ph]].

\bibitem{Branchina:2014usa}
  V.~Branchina, E.~Messina and A.~Platania,
  JHEP {\bf 1409} (2014) 182
  [arXiv:1407.4112 [hep-ph]].

\bibitem{Branchina:2014rva}
  V.~Branchina, E.~Messina and M.~Sher,
  arXiv:1408.5302 [hep-ph].
  
  
\bibitem{Bezrukov:2010jz}
  F.~Bezrukov, A.~Magnin, M.~Shaposhnikov and S.~Sibiryakov,
  JHEP {\bf 1101} (2011) 016
  [arXiv:1008.5157 [hep-ph]].

\bibitem{Bezrukov:2014ipa}
  F.~Bezrukov, J.~Rubio and M.~Shaposhnikov,
  arXiv:1412.3811 [hep-ph].
  
\bibitem{Longhitano:1980iz}
  A.~C.~Longhitano,
  Phys.\ Rev.\ D {\bf 22} (1980) 1166.
  
  
\bibitem{Bezrukov:2009db}
  F.~Bezrukov and M.~Shaposhnikov,
  JHEP {\bf 0907} (2009) 089
  [arXiv:0904.1537 [hep-ph]].

  
\bibitem{Bezrukov:2014bra}
  F.~Bezrukov and M.~Shaposhnikov,
  Phys.\ Lett.\ B {\bf 734} (2014) 249
  [arXiv:1403.6078 [hep-ph]].

\bibitem{Hamada:2014iga}
  Y.~Hamada, H.~Kawai, K.~y.~Oda and S.~C.~Park,
  Phys.\ Rev.\ Lett.\  {\bf 112} (2014) 241301
  [arXiv:1403.5043 [hep-ph]].

 \bibitem{GarciaBellido:2008ab}
  J.~Garcia-Bellido, D.~G.~Figueroa and J.~Rubio,
  Phys.\ Rev.\ D {\bf 79} (2009) 063531
  [arXiv:0812.4624 [hep-ph]].
 

 \bibitem{Bezrukov:2008ut}
  F.~Bezrukov, D.~Gorbunov and M.~Shaposhnikov,
  JCAP {\bf 0906} (2009) 029
  [arXiv:0812.3622 [hep-ph]].
  
\bibitem{Anderson:1990aa}
  G.~W.~Anderson,
  Phys.\ Lett.\ B {\bf 243} (1990) 265.
  
\bibitem{Arnold:1991cv}
  P.~B.~Arnold and S.~Vokos,
  Phys.\ Rev.\ D {\bf 44} (1991) 3620.
  
\bibitem{Espinosa:1995se}
  J.~R.~Espinosa and M.~Quiros,
  Phys.\ Lett.\ B {\bf 353} (1995) 257
  [hep-ph/9504241].
  
\bibitem{Espinosa:2007qp}
  J.~R.~Espinosa, G.~F.~Giudice and A.~Riotto,
  JCAP {\bf 0805} (2008) 002
  [arXiv:0710.2484 [hep-ph]].


  

\end{thebibliography}
\end{document}